\documentclass[lettersize,journal]{IEEEtran}
\usepackage{amsmath,amsfonts}
\usepackage{amssymb}
\usepackage{algorithmic}
\usepackage{makecell}
\usepackage{booktabs}
\usepackage{array}
\usepackage[caption=false,font=normalsize,labelfont=sf,textfont=sf]{subfig}
\usepackage{textcomp}
\usepackage{caption}
\usepackage{stfloats}
\usepackage{multirow} 
\usepackage{url}
\usepackage[numbers,sort&compress]{natbib}
\usepackage[breaklinks,colorlinks]{hyperref}
\usepackage{verbatim}
\usepackage{graphicx}
\def\BibTeX{{\rm B\kern-.05em{\sc i\kern-.025em b}\kern-.08em
    T\kern-.1667em\lower.7ex\hbox{E}\kern-.125emX}}
\usepackage{balance}
\usepackage{colortbl}
\usepackage{xcolor}

\begin{document}

\title{A Unified Framework for Modality-Agnostic Deepfakes Detection}

\author{Cai Yu, Peng Chen, Jiahe Tian, Jin Liu, Jiao Dai, Xi Wang, Yesheng Chai, Shan Jia, Siwei Lyu, \IEEEmembership{Fellow, IEEE}, Jizhong Han, \IEEEmembership{Member, IEEE}
\thanks{ C. Yu, J. Tian, and J. Liu are with the Institute of Information Engineering, Chinese Academy of Sciences, Beijing, China, and
also with the School of Cyber Security, University of Chinese Academy
of Sciences, Beijing, China. E-mail: (caiyu, tianjiahe,
liujin)@iie.ac.cn.} 
\thanks{P. Chen is with the RealAI Inc, Beijing, China. E-mail: peng.chen@realai.ai.}\\
\thanks{J. Dai, X. Wang, Y. Chai, and J. Han are with the Institute of Information Engineering, Chinese Academy of Sciences, Beijing,
China. E-mail: (daijiao, wangxi1, chaiyesheng, 
hanjizhong)@iie.ac.cn.}
\thanks{S. Jia, and S. Lyu are with the Department of Computer
Science and Engineering, University at Buffalo, State University of
New York, NY, USA. E-mail: (shanjia, siweilyu)@buffalo.edu.}}

% \markboth{Journal of \LaTeX\ Class Files,~Vol.~18, No.~9, September~2020}%
% {How to Use the IEEEtran \LaTeX \ Templates}

\maketitle

\begin{abstract}

As AI-generated content (AIGC) thrives, deepfakes have expanded from single-modality falsification to cross-modal fake content creation, where either audio or visual components can be manipulated. While using two unimodal detectors can detect audio-visual deepfakes, cross-modal forgery clues could be overlooked. Existing multimodal deepfake detection methods typically establish correspondence between the audio and visual modalities for binary real/fake classification, and require the co-occurrence of both modalities. However, in real-world multi-modal applications, missing modality scenarios may occur where either modality is unavailable. In such cases, audio-visual detection methods are less practical than two independent unimodal methods. Consequently, the detector can not always obtain the number or type of manipulated modalities beforehand, necessitating a fake-modality-agnostic audio-visual detector. In this work, we introduce a comprehensive framework that is agnostic to fake modalities, which facilitates the identification of multimodal deepfakes and handles situations with missing modalities, regardless of the manipulations embedded in audio, video, or even cross-modal forms. To enhance the modeling of cross-modal forgery clues, we employ audio-visual speech recognition (AVSR) as a preliminary task. This efficiently extracts speech correlations across modalities, a feature challenging for deepfakes to replicate. Additionally, we propose a dual-label detection approach that follows the structure of AVSR to support the independent detection of each modality. Extensive experiments on three audio-visual datasets show that our scheme outperforms state-of-the-art detection methods with promising performance on modality-agnostic audio/video deepfakes. 
\end{abstract}

\begin{IEEEkeywords}
Multimedia forensics, Multimodal learning, Video Forgery detection.
\end{IEEEkeywords}

\section{Introduction}
\IEEEPARstart{A}{s} artificial intelligence-generated content (AIGC) technologies continue to advance, deepfakes are evolving to become increasingly complex and diverse in form. 

Traditional deepfakes mainly generate fake faces or voices tied to a specific person within a single modality, either through visual deepfake generation techniques~\cite{korshunova2017fast,thies2016face2face,siarohin2019first} or audio synthesis methods~\cite{ping2018clarinet,kameoka2018non}. For instance, a deepfake video falsely depicting New Zealand Prime Minister Jacinda Ardern smoking crack cocaine circulated widely on social media\footnote{\url{https://www.youtube.com/shorts/j0v4UMnHn1M}}. In another example, AI-generated audio mimicking Donald Trump was used to scam Trump voters\footnote{\url{https://www.tiktok.com/t/ZT8MWGQ4g/}}. These instances underscore the growing sophistication and potential impact of deepfake technologies. However, the landscape of deepfake technology is rapidly evolving. Cutting-edge deepfake methods like Wav2Lip~\cite{prajwal2020lip} and VideoReTalking~\cite{cheng2022videoretalking} can even achieve cross-modal forgery by driving audio to generate precise lip-sync videos. To date, malicious attackers have already been capable of combining the above technologies to create multimodal forgeries in which both the face and voice of the target individual can be seamlessly manipulated.
 \begin{figure}[t]
 \centering
\includegraphics[width=1.0\columnwidth]{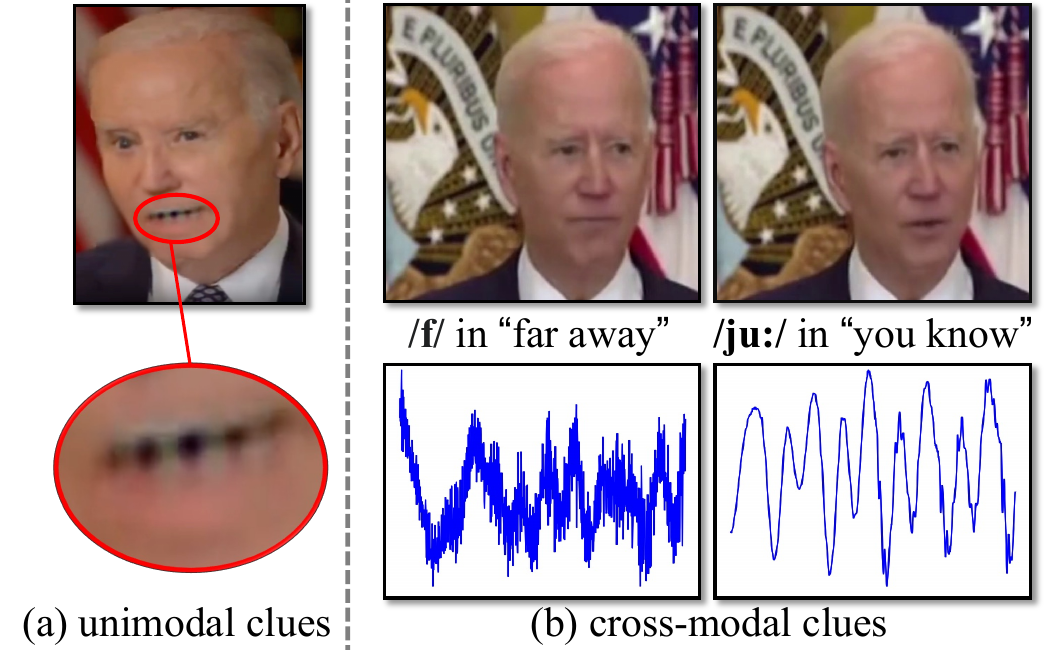} % Reduce the figure size so that it is slightly narrower than the column.
 \caption{Forgery patterns in deepfakes\protect\footnotemark. (a) Visual artifacts within a single modality: Biden's teeth exhibit an irregular shape. (b) Cross-modal speech mismatch in deepfakes. The phoneme-level audio waveform corresponds to the duration of the frame shown above. Both images reveal a speech mismatch between lip motions and audio sounds. On the left, Biden's mouth is in a closed state before pronouncing the ``f" sound and has not yet produced a complete syllable. However, the audio image displays a complex frequency wave that should only appear during active pronunciation. On the right, Biden is uttering the vowels in ``you", but the sound waves show rather simple frequency patterns, which are typically indicative of panting or background noise rather than speech.}
 \label{fig:Biden}
\end{figure}
\footnotetext{videos of these examples are cited from \url{https://deepfakes.lol/}, and \url{https://www.tiktok.com/t/ZT8M7GmQg/}}As video examples in Figure \ref{fig:Biden} show, the fake `Biden' in these videos is uttering nonsense in a voice that exactly mimics his own. Compared to unimodal deepfakes, multimodal deepfakes, owing to their multi-faceted forgeries, could be more realistic and thus much harder to discern. 

Consequently, deepfakes have evolved to include not only audio and visual but also cross-modal forms, thereby exacerbating their detrimental impact accordingly. This evolution raises new challenges for detection methods, making the fight against deepfakes even more complex: Deepfake content can now appear in various forms across social media (audio-only, video-only, or audio-visual pairs), and a single detector may not always have access to the range or type of manipulated modalities beforehand. While employing two separate unimodal detectors seems like a viable approach, the cross-modal forgery clues generated might be overlooked. The comparison in Figure \ref{fig:Biden} shows that the fake frames in (b) do not exhibit visual artifacts as obvious as those in (a). Yet, when paired with audio, these frames reveal a phoneme-level mismatch in speech. This observation implies that when no readily apparent forgery patterns are detectable within a single modality, employing a cross-modal detector could be more effective in capturing forgery patterns across multiple modalities.
 Considering the significance of multimedia content security, it is crucial to develop a unified, multimodal framework that can effectively defend against emerging deepfakes, irrespective of the number or type of modalities involved.
% \footnotetext[3]{\url{https://deepfakes.lol/}}
% \footnotetext[4]{\url{https://www.tiktok.com/t/ZT8M7GmQg/}}

\begin{figure*}[!htb]
\centering
  \includegraphics[width=\textwidth]{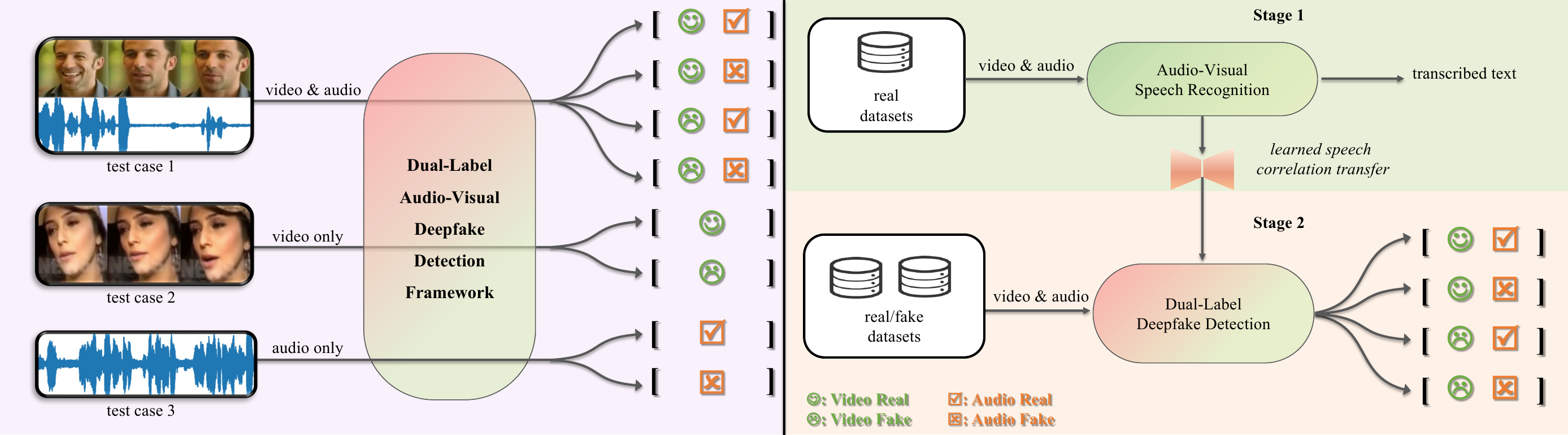}
  \caption{Modality-Agnostic Audio-Visual Deepfake Detection: The figure on the left shows three modality-agnostic detection scenarios that our detector supports. This detector enables independent detection of forgery in each modality and can handle scenarios where either modality is unavailable. The figure on the right illustrates the general scheme of our method, which leverages the speech correlation across modalities via the AVSR task to perform dual-label detection.}
  %\Description{Enjoying the baseball game from the third-baseseats. Ichiro Suzuki preparing to bat.}
  \label{fig:teaser}
\end{figure*}

Nowadays, a few emerging audio-visual deepfake detection methods~\cite{chugh2020not, gu2020deepfake, mittal2020emotions} have achieved promising results. For example, EmoForen~\cite{mittal2020emotions} detects deepfakes via emotional inconsistency. These methods typically treat audio as a reference signal to assess the level of audio-visual consistency, ultimately generating a single binary real/fake label for the entire video. 
 
More recent methods, such as JointAV~\cite{zhou2021joint} and AVoiD-DF~\cite{yang2023avoid}, attempt to leverage cross-attention to learn potential cross-modal dependence. However, more effective perceptual cues remain underutilized. Besides, all the abovementioned methods necessitate the presence of audio and video in pairs. However, in the real world, deepfakes do not always appear in audio-visual pairs. There are detection scenarios where either the audio or visual component of the media is missing or unavailable. In such modality-agnostic situations, relying solely on audio-visual detection methods may be less practical than employing two separate unimodal detection methods. Nevertheless, audio-visual detection remains vital for identifying cross-modal forgery clues. This raises a critical question: Is there a ``best of both worlds" approach for an audio-visual detection framework that can effectively leverage cross-modal forgery clues while still maintaining robust performance in modality-agnostic scenarios? 

In this work, we propose a unified framework for audio-visual deepfake detection that targets modality-agnostic scenarios. First, to effectively capture cross-modal forgery clues, we introduce Audio-Visual Speech
Recognition (AVSR) as a preliminary task (as illustrated on the right of
Figure \ref{fig:teaser}). This practice is motivated by the observation that natural phonation processes exhibit consistent dynamic patterns between audio and visual modalities. Conversely, such subtle, speech-related correlations may be challenging for deepfakes to replicate due to their unnatural origins.
By pretraining on a large number of real videos, the AVSR model learns to map both audio and visual signals to the same set of spoken characters. This makes it well-suited for extracting natural speech correlations across modalities. By transferring this speech correlation into our framework, the modeling of cross-modal forgery clues can be enhanced. Customized for deepfake detection, a Modality Compensation Adapter is inserted into the AVSR model to alleviate modality imbalance in model decisions. Second, to support the independent detection of each modality, we propose a dual-label detection approach where independent labels are assigned for each modality so that even if one modality is missing, the detection of the other will not be affected (as illustrated on the left of
Figure \ref{fig:teaser}). Based on this insight, a Dual-Label Classifier is designed and established with a Fake Composition Detector to extract forgery intensity and a Temporal Aggregation Module to capture temporal inconsistency. The Dual-Label detection approach establishes a parallel structure between each modality, which aligns with the AVSR structure, thereby maximizing AVSR's advantages. The unified structure formed by the combination of AVSR and Dual-Label Classifier guarantees the ability to effectively combat modality-agnostic deepfakes.
The contributions of this work are summarized as follows:

\begin{itemize}
    \item We propose a unified audio-visual deepfake detection framework, incorporating AVSR as a preliminary task to model cross-modality correlation. Tailored for multimodal deepfake detection, we design a Modality Compensation Adapter within the AVSR architecture to effectively balance the visual and audio modalities. 
    
    \item We advocate for a modality-agnostic approach to handle various deepfakes and reformulate audio-visual deepfake detection into a dual-label framework. A Dual-Label Classifier is developed to independently predict the authenticity of each modality. A Temporal Aggregation Module and a Fake Composition Detector are further integrated to enhance the classifier. To the best of our knowledge, this is the first work to enable detection of both multi-modal and single modality deepfakes.

   \item Extensive experiments on three datasets demonstrate the superiority and flexibility of our framework in detecting modality-agnostic deepfakes.  
\end{itemize}

\section{Related Work}
Extensive research has been dedicated to detecting deepfakes in multimedia content. According to the modality involved, methods for deepfake detection can be broadly divided into three categories as follows.

\begin{table*}[!t]
\centering
\caption{Comparison of various audio-visual deepfake detection methods. Methods are evaluated based on the approach of modeling cross-modal forgery patterns (Forgery Pattern Modeling), the type of classification approach used (Classification Approach), their capability to specify which modality has been manipulated (Forgery Modality Identification), and whether they can detect modality-agnostic deepfakes when a modality is missing (Support for Missing Modality).}

\begin{tabular}{lcccccc}
\toprule
\multirow{2}{*}{\makecell{Methods}} &\multirow{2}{*}{\makecell{Year}} & \multirow{2}{*}{\makecell{Forgery Pattern\\ Modeling }}  & \multirow{2}{*}{\makecell{Classification \\ Approach }} & \multirow{2}{*}{\makecell{Forgery Modality \\ Identification }} & \multirow{2}{*}{\makecell{Support for \\Missing Modality}} \\
%\cmidrule(lr){}
 \\
\midrule
MDS~\cite{chugh2020not} &2020 &Modality Dissonance Score  & Binary &$\times$ & $\times$ \\
EmoForen~\cite{mittal2020emotions} &2020 &Emotion Consistency  & Binary &$\times$ & $\times$ \\
JointAV~\cite{zhou2021joint}  &2021 &Inter-modal Attention  & Triple-Classifier &$\checkmark $ & $\times$ \\
VFD~\cite{cheng2022voice} &2022  &Idengtity Similarity   &Binary &$\times  $ & $\times$ \\
RealForensics~\cite{haliassos2022leveraging} &2022  &Cross-modal Self-supervised Learning  & Binary &$\times$ & $\times$ \\
BA-TFD~\cite{cai2022you} &2022  &Temporal Synchronization   &Binary &$\times  $ & $\times$ \\
AVoiD-DF~\cite{yang2023avoid} &2023 &Multi-modal Joint-decoder   &Binary &$\times  $ & $\times$ \\
AVFakeNet~\cite{ilyas2023avfakenet} &2023 &Swin Transformer & Dual-Classifier &$\checkmark $ & $\times$ \\

AVAD~\cite{feng2023self} &2023 & Autoregressive Synchronization  & Binary &$\times$ & $\times$ \\
PVASS~\cite{yu2023pvass} &2023 &Self-supervied Visual-audio Alignment  & Binary &$\times$ & $\times$ \\
Ours &2023 &Speech Correlation   & Dual-Label &$\checkmark $ & $\checkmark$ \\
% Method A & \checkmark & \checkmark & $\times$ & \checkmark & $\times$  
% Method B & \checkmark & \checkmark & \checkmark & $\times$ & \checkmark\\
\bottomrule
\end{tabular}
\label{classfiy_compare}
\end{table*}

\subsection{Visual Modality Deepfake Detection} 
These methods perform image-level or video-level real/fake prediction. Image-level methods focus on image or intra-frame artifacts. Face X-ray \cite{li2020face} and SBIs \cite{shiohara2022detecting} target forgery artifacts of blending boundaries by simulating deepfake images. PEL \cite{gu2022exploiting}, PADD~\cite{yu2022focus}, and MaDD \cite{zhao2021multi} exploit fine-grained clues to detect face forgery. As to video-level methods, previous works leverage biometric signals to detect deepfake videos, such as facial action patterns \cite{agarwal2019protecting,sun2021improving} and eye blinking \cite{li2018eye}. Another branch of video-level works utilizes temporal modeling~\cite{gu2022delving,zhang2021detecting,gu2021spatiotemporal} for classification since manipulation in fake videos usually causes temporal inconsistency. Arguing that the above tangible clues may make models overfit to low-level manipulation-specific artifacts, LipForens \cite{haliassos2021lips} proposes to catch high-level semantic inconsistency via a lipreading network. Regarding the utilization of high-level information, LipForens is highly relevant to our work. However, unlike our proposal, the utility of audio has been neglected.

\subsection{Auditory Modality Deepfake Detection} 
Audio deepfake detection mainly counters the voice spoofing synthesized by texts-to-speech (TTS) or voice conversion (VC) techniques. Some works \cite{martin2022vicomtech,xie2021siamese,wang2021investigating} apply pretrained speech representation models, such as wav2vec2 \cite{baevski2020wav2vec}, to detect spoofed audio. Other works leverage classic speech tasks to improve audio deepfake detection; for example, DAD-SV~\cite{pianese2022deepfake} boils down deepfake detection into a speaker verification problem. These works also render a high plausibility for us to introduce speech recognition to detect multimodal deepfakes.
%and DAD-SV~\cite{pianese2022deepfake} 

\subsection{Audio-Visual Deepfake Detection} 
%,yu2023pvass,liu2023mcl
There exist a few methods involving both visual and auditory modalities. However, most of them intrinsically measure the degree of audio-visual correspondence~\cite{chugh2020not,mittal2020emotions,hosler2021deepfakes} or treat audio signals as reference \cite{haliassos2022leveraging} for video so that they can only assign a binary real/fake label for the entire video. For example, MDS
\cite{chugh2020not} measures the audio-visual dissonance in a video via the Modality Dissonance Score. EmoForen~\cite{mittal2020emotions} leverage emotion clues to discern fake videos. %Nevertheless, these approaches barely achieve synergistic detection of both visual and auditory manipulation. 
Nevertheless, these methods can only determine whether audio and visual components are consistent or not; they cannot specifically predict which modality has been manipulated. JointAV \cite{zhou2021joint} and AVFakeNet~\cite{ilyas2023avfakenet} have attempted to detect manipulation in each modality using a multi-classifier approach. However, they primarily rely on the attention mechanism to capture cross-modal dependencies of the audio-visual pairs, while the high-level perceptual information such as speech correlation is underutilized. Moreover, none of the aforementioned methods have discussed how their multimodal detectors perform when encountering scenarios with missing modalities. %A detailed comparison of current audio-visual deepfake detection methods is shown in Table \ref{classfiy_compare}.
For a detailed comparison of current audio-visual deepfake detection methods, refer to Table \ref{classfiy_compare}.

\section{How AVSR Benefits Deepfake Detection}
In this section, we aim to provide a comprehensive explanation of our motivation behind introducing AVSR as the preliminary task, focusing specifically on two aspects.
\subsection{Biological Perspective}
 Human perception of speech involves both audition and vision. The cortical correlates of seen speech suggest that auditory processing of speech is affected by vision at both neurological and perceptual levels \cite{shi2021learning}. Biological studies have shown that for speech perception, the visual channel and auditory channel share similar dynamic patterns \cite{campbell2008processing}. Thus, we seek to leverage this kind of high-level audio-visual correlation, or, more precisely, audio-visual speech correlation \cite{almajai2007maximising}, to identify real/fake audio-visual media. The rationale is that the correlation between audio and visual speech features is an intrinsic characteristic of the natural phonation process, which can be absent in deepfake videos due to the artificial synthesis process. In deepfakes, the visual and audio components are typically initially separated, allowing either of them to be synthesized individually before being merged together. This will inevitably lead to abnormal audio-visual correlations. A study by \cite{9151013} has identified phoneme-viseme mismatches in deepfakes when ``M", ``B", and ``P” are pronounced. Our observations in Figure \ref{fig:Biden} (b) reveal a similar disruption in speech correlation across modalities. As the feature quality of audio-visual correlation can be reflected by AVSR~\cite{almajai2007maximising}, we choose it as the preliminary task in our framework to enhance the modeling of cross-modal forgery clues.
\begin{figure*}[!t]
  \centering
  \includegraphics[width=1.0\textwidth{},keepaspectratio]{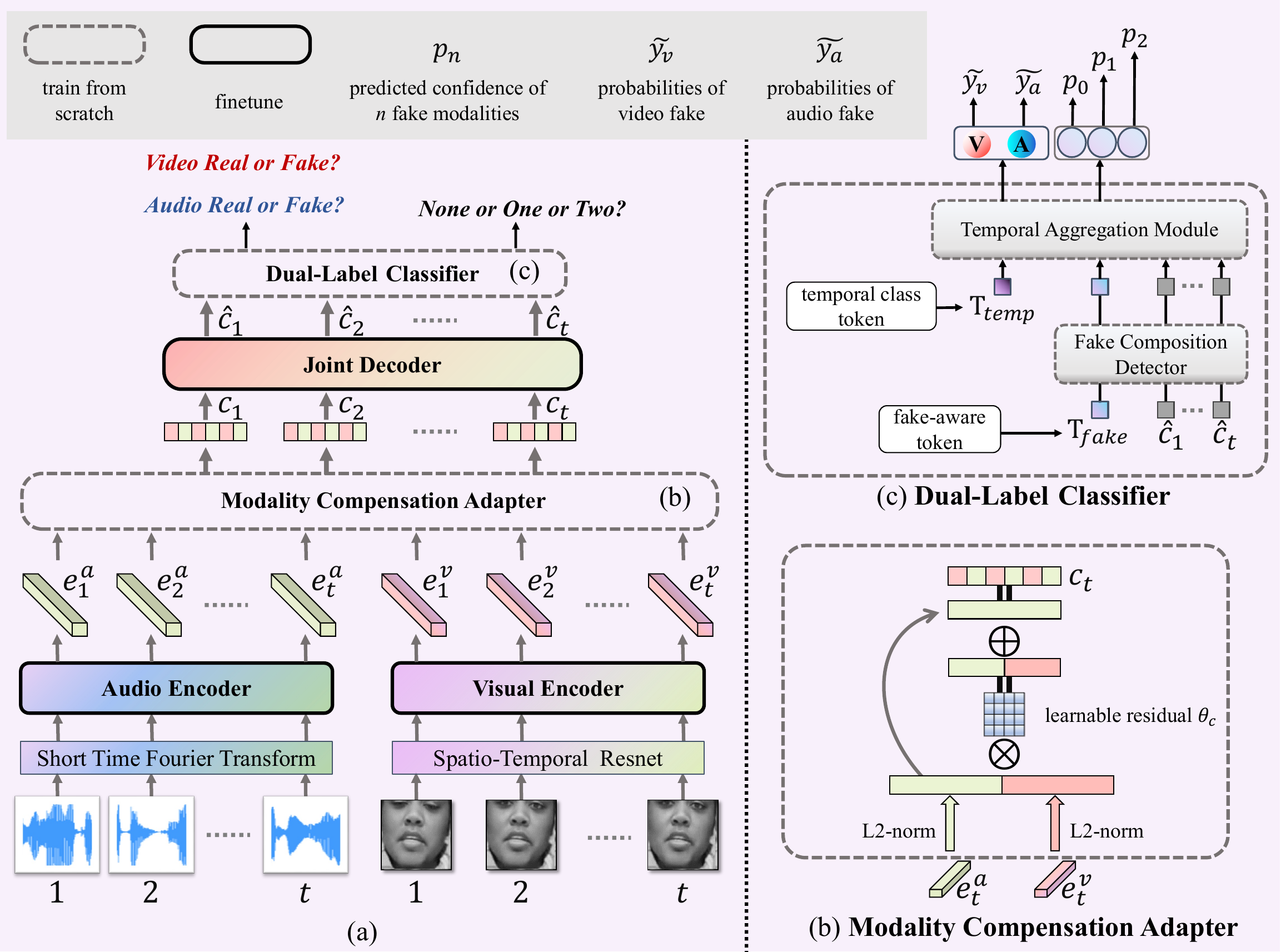}
  \caption{Demonstration of learning scheme of dual-label deepfake detection. (a) the overview of the proposed framework. The two encoders and joint decoder, which are pretrained on AVSR, are finetuned to model speech correlation across modalities. (b) Modality Compensation Adapter  (MCA) is embedded into the encoder-decoder to prevent the network's over-reliance on a certain modality. (c) Dual-Label Classifier (DLC) is attached at the backend of the model to perform both audio and visual manipulation detection at once.}\label{fig:overview}
\end{figure*}
 \subsection{ Applications Perspective}
 % 1. cross-modal correlation weak, 2.AVSR 2018 attempts to enable missing modality. for example, deepfake  same requirements. 
% 
In real-world multimodal applications, it is common to encounter situations where either the audio or visual component is missing or unavailable~\cite{ma2021smil}. This also applies to the field of audio-visual deepfake detection. For our framework, we aim to strike a balance in cross-modal dependencies—sufficiently strong to mutually enhance the detection of each modality, yet not overly strong as to compromise performance when a specific modality is missing. The application scenario of AVSR aligns well with this objective, offering a moderate level of cross-modal dependency. 
Before the advent of AVSR, both ASR (Audio Speech Recognition)~\cite{baevski2020wav2vec} and VSR (Visual Speech Recognition)~\cite{7952625} have already exhibited effective performance. The significance of AVSR primarily lies in its ability to compensate for the limitations of these two tasks, such as learning more robust speech representations by leveraging visual signals in noisy speech environments or using speech signals to assist lip-reading models when mouth shapes look similar (e.g., differentiating ``b" and ``p" phoneme could be better with audio signal combined~\cite{afouras2018deep}). Some existing AVSR works have also demonstrated the ability to perform signal-to-character mapping within each modality independently, even when the other modality is absent~\cite{8682566,9004036}. This suggests that the cross-modal speech-related dependency in AVSR not only improves speech recognition in each unimodal model but also maintains semantic integrity within each modality when one is missing. This makes AVSR well-suited for detecting audio-visual deepfakes in complex scenarios, aligning it closely with our modality-agnostic objectives.

\section{Method}
Given that the probability of forgery for each modality in a video is independent, and either modality may or may not be present, we contend that traditional binary classification methods are not ideal and fall short in handling modality-agnostic scenarios. Our goal is to holistically identify any forged modalities in a video, irrespective of their number or type. To address this, we formulate the deepfake detection task as a dual-label detection approach. This approach is designed to concurrently and independently predict the authenticity of both modalities within a single task. The proposed framework consists of two stages. At the first stage, an encoder-decoder architecture is pretrained on the task of AVSR to extract desired speech correlation. The second stage performs deepfake detection based on acquired speech knowledge. A detailed description of the proposed framework is illustrated in the following subsections.

\subsection{Stage 1: Audio-Visual Speech Recognition}

Given a speech recognition dataset $D_{sr}=\left\{a_{t}^{i}, v_{t}^{i}, y_{t}^{i} \right\}_{i=1}^{N_{sr}}$ of size $N_{sr}$, where $t = 1,...,T$ denotes time step of input sequence length and $a_{t}^{i},v_{t}^{i}$ are the corresponding audio segment and video frame of each time step. For speech recognition task, each time step contains a ground-truth label (word or character) $y_{t}^{i}$. In the setting of this framework, the network produces frame-wise character probabilities $\tilde{y}_{t}^{i} \in\{1,...,C\}$, where $C$ is the character class. In our implementation, the set $\tilde{y}_{t}^{i} \in\{1,...,C\}$ consists of 40 phoneme characters.

Speech recognition is a sequence model, which comprises two unimodal encoders $\theta_{a},\theta_{v}$, a joint decoder $\theta_{j}$ and a multi-class classifier $\theta_{c}$. It first extracts unimodal embedding through each encoder, referred to as, $e_{t}^a={\theta_{a}}(a_{t}),e_{t}^v={\theta_{v}}(v_{t})$. The final character prediction is obtained by passing the concatenation of two modal embeddings through the decoder and the classifier: $\tilde{y}_{t} = \theta_c({\theta_{j}}\left(e_{t}^a, e_{t}^v\right))$. Following \cite{afouras2018deep}, the speech recognition network is trained with CTC (Connectionist Temporal Classification) loss \cite{graves2006connectionist}, 
\begin{equation} \frac{1}{N_{sr}} \sum_{i=1}^{N_{sr}} \mathcal{L}_{\mathrm{CTC}}\left(\tilde{y}_{t}^{i}, y_{t}^{i} ; \theta_{a}, \theta_{v}, \theta_{j},\theta_{c}\right)\end{equation}

\subsection{Stage 2: Audio-Visual Deepfake Detection}

The process of dual-label deepfake detection is illustrated in Figure \ref{fig:overview}a. 
The weights of encoders and the decoder are transferred from that ($\theta_{a}, \theta_{v}, \theta_{j}$) of speech recognition. A specially designed Modality Compensation Adapter  $\theta_{m}$ is inserted into the encoder-decoder and trained from scratch. To perform dual-label detection, we design a dual-label classifier at the end of the network. The design of each component is further detailed in the subsequent subsection.
\subsubsection{\textbf{Problem Formulation}}
%Given a deepfake detection dataset $D_{df}=\left\{a_{t}^{j}, v_{t}^{j}, y_{m}^{j} \right\}_{j=1}^{N_{df}}$, where $y^{m} \in\{0,1\}$, $m \in\{a,v\}$,where $m$ is the modality (i.e. video $a$ or audio $v$),
 Given a deepfake detection dataset $D_{df}=\left\{a_{t}^{j}, v_{t}^{j}, y_{m}^{j} \right\}_{j=1}^{N_{df}}$, where $y_{m} \in\{0,1\}$ is a time-independent label for the entire video, and $m \in\{a,v\}$ denotes the modality dimension (\emph{i.e.}, audio $a$ or video $v$), the goal of our framework is to predict the $y_{m}^{j}$ based on the input $a_{t}^{j}$ and/or $v_{t}^{j}$. Compared with a conventional binary prediction for the entire video, where $y^i=0$ for real or $y^i=1$ for fake, the dual-label is reflected in that each dimension of the label specifies the possibility of whether each modality has been forged. For example, $y^i=[0,1]$ denotes the audio is real and 
video is fake, and $y^i=[1,1]$ denotes that both audio and video are fake.

\subsubsection{\textbf{Modality Compensation Adapter}}
\label{methodMCA}
This module is specifically designed for the characteristics of multimodal deepfake detection, aiming at enhancing the adaptability of pre-learned speech correlations to our task. As demonstrated in the recent speech recognition literature, AVSR models can relate audio modality to lexical prediction more effortlessly than the visual modality \cite{shi2021learning,afouras2018deep,ma2021end}. This might cause the audio modality to dominate the model decisions. In terms of deepfake detection, we believe that the differences between reals and fakes may vary across modalities, which could also lead to the model becoming overly reliant on a specific modality. To verify if there is a modality dominance issue in our task, we perform an exploratory baseline experiment of dual-label deepfake detection on randomly sampled videos from the DFDC dataset~\cite{dolhansky2019deepfake}. 
\begin{table}[h]
\centering
\caption{An exploratory experiment on detection of each modality in dual-label manner.}
\begin{tabular}{l|l|l}
\hline
       & Recall & \multicolumn{1}{c}{F1} \\ \hline
Visual & 76.92 & 77.39                   \\ \hline
Audio  & 57.81 & 71.49                   \\ \hline
\end{tabular}

\label{tab:oberving}
\end{table}

As shown in Table~\ref{tab:oberving}, the performance of audio detection is significantly worse than that of video detection. Hence, it is reasonable to speculate that there appears to be some level of visual dominance in the model decisions. One potential reason is that visual data is much richer than audio data in amount and diversity. Although a straightforward solution is to collect more audio samples, we think this manner requires much time and effort and is not flexible. %Although this can be alleviated by using more fake audio samples, we think this manner is palliative.

To address the above issue of modality dominance, we propose a Modality Compensation Adapter  (Figure \ref{fig:overview}b). Here, we only formalize the compensation for the audio modality below, and the same operation can be applied to the video counterpart.
\begin{equation} 
c_{t}=e_{t}^{a}+\theta_{c}\left(Norm_{L2}\left(e_{t}^{a}\right),Norm_{L2}\left(e_{t}^{v}\right)\right)
 \end{equation} 

 where $e_{t}^{a}$, $e_{t}^{v}$ are unimodal embeddings extracted by encoders and $c_{t}$ denotes the compensated embeddings. Here we compensate audio modality by concatenating $L2$ normalization of each unimodal embedding and passing them through a learnable residual $\theta_{c}$.
 We adopt audio compensation in our framework considering its contribution to balancing the two modalities and thus yielding more reasonable and preferable results that align with our speculation (as verified by the ablation study in Section \ref{subsection MC}).%because, compared to video compensation, it yields more reasonable and preferable results that align with our speculation (as verified by the ablation study in section \ref{subsection MC}). 

\subsubsection{\textbf{Dual-Label Classifier}}

To achieve the goal of synergistic detection of both visual and auditory manipulation, we design a dual-label classifier that is composed of a Fake Composition Detector (FCD) and a Temporal Aggregation Module (TAM). This design is implemented based on two considerations. First, in the speech recognition architecture, it performs frame-wise prediction for each time step, whereby only frame-level outputs are served. However, for deepfake detection, frame-level output is insufficient to capture temporal jitter between consecutive frames. Straightforward conduction of a temporal-pooling operation might average these subtle clues, thus an interactive video-level information aggregation is needed to capture temporal inconsistency. Second, the real/fake discrepancy varies from different modalities. To bridge the forgery intensity gap across modalities and provide a generalizable representation, we impose an extra supervision from the given label to denote the forgery intensity, \emph{i.e.}, the number of fake modalities in a given video. This practice is inspired by MlTr \cite{cheng2021mltr}, a successful work in multi-label classification because dual-label deepfake detection shares a similar concept with multi-label classification tasks in terms of the classification paradigm. In their work, they claim that this reinforced supervision can force the model to extract the common features of the same class and learn a more robust projection between features and labels, while our implementation aims at endowing a constraint of the same intensity to the audio or video manipulation in an explicit way. Specifically, despite their forgery intensity difference, in the case of only one modality has been manipulated in a certain video, whether it is visual, or audio, the intense label is assigned as 1. We believe that this design can provide a more generalizable representation, which is validated in the experiment section.

Based on the above consideration, we propose to implement these designs in our dual-label classifier via two transformer-like \cite{vaswani2017attention} modules. FCD serves for forgery intensity modeling and TAM for capturing temporal inconsistency. As shown in Figure \ref{fig:overview}c, leveraging the flexibility of transformer tokens, we prepend two randomly initialized tokens at the head of the feature sequence: a fake-aware token $\mathbf{{T}}_{\text {fake }}$, and a temporal class token $\mathbf{{T}}_{\text {temp }} $, respectively. Formally, we redefine the compensated embeddings that have passed through the joint decoder as \( \hat{c_t} \). The dimensions of the two tokens are the same as those of \( \hat{c_t} \). We then proceed to perform dual-label classification as follows:

\begin{equation}
\left.\mathbf{Seq}_{\text {temp }}=FCD\left(\ { Concat(\mathbf{{T}}_{\text {fake }}; \hat{c_{1}} ; \hat{c_{2}} \ldots \hat{c_{T}}}\right)\right)\\
\end{equation}
\begin{equation}
\left.\mathbf{Seq}=TAM\left(\ { Concat (\mathbf{{T}}_{\text {temp }}  } ,\mathbf{Seq}_{\text {temp }}\right)\right)\\
\end{equation}
\begin{equation}
\tilde{y}_m^j=M L P(\mathbf{Seq}[0]) ,
     \tilde{p}^j=MLP\operatorname(\mathbf{Seq}[1])\\
\end{equation}
% \begin{equation}
% \mathbf{\hat{Token}}_{\text {fake }},\mathbf{\hat{Token}}_{\text {time }} =z[0], z[1]
% \end{equation}

where \( \mathbf{Seq} \) denotes the resultant output after FCD and TAM. \( \mathbf{Seq}[0] \) and \( \mathbf{Seq}[1] \) represent the corresponding outputs of the temporal class token \( \mathbf{T}_{\text{temp}} \) and the fake-aware token \( \mathbf{T}_{\text{fake}} \), respectively, which are then processed by a multi-layer perceptron (MLP) to output the final predictions. The difference is that the temporal class token predicts specific forgery modalities $\tilde{y}_{m}^{j}$ and is optimized by binary cross-entropy loss, 
\begin{equation}  L_{bce}=\frac{1}{N_{df}} \sum_{j=1}^{N_{df}} \mathcal{L}_{\mathrm{BCE}}\left(\tilde{y}_{m}^{j}, y_{m}^{j} ; \theta_{a}, \theta_{v}, \theta_{m} ,  \theta_{j},\theta_{d}\right)
\end{equation}
while the fake-aware token outputs a prediction score for the number of fake modalities. Let \( \tilde{p}^j \) and \( {p}^j \) represent the prediction and ground truth of the fake-aware token, respectively. The ground truth is given by \( p^j = \sum_{m \in \{a, v\}} y_{m}^{j} \in \{0, 1, 2\}\) to quantifies the number of fake modalities. The prediction is subsequently optimized using the cross-entropy loss,

\begin{equation} L_{ce}=\frac{1}{N_{df}} \sum_{j=1}^{N_{df}} \mathcal{L}_{\mathrm{CE}}\left( \tilde{p}^j,p^{j}; \theta_{a}, \theta_{v}, \theta_{m} ,  \theta_{j},\theta_{d}\right)
\end{equation}

Eventually, for the final framework, the overall learning objective can be defined as:
\begin{equation}
L= L_{bce}+ L_{ce}
\end{equation}
%where  $\lambda_{f}$ is a weight factor, which is set to 1.

%\subsection {Pretraining of AVSR}

 %

\section{EXPERIMENTS AND RESULTS}
\subsection {Preprocessing}
\subsubsection{\textbf{Audio Stream}}
The audio signals are all resampled at a 16kHz sample rate and applied Short Time Fourier Transform (STFT) to obtain 321-dimensional spectrograms, with a 40ms window and 10ms hop-length. Since the visual signals are sampled at 25 fps per video, each visual frame corresponds to 4 acoustic frames. Following AVSR, the acoustic frames are padded to make the input length a multiple of 4. Every 4 acoustic frames are concatenated and passed through a 1D convolutional layer with stride 4 to align input length with visual frames.

\subsubsection{\textbf{Video Stream}}
After sampling the video at 25fps, we perform face tracking using the S3FD face detector \cite{zhang2017s3fd} for the video datasets that are not face-centered. For face-centered video frames, we resize it into a $224\times224$ and crop out a $112 \times 112$ patch that covers mouth region. Following Deep-AVSR, given a video of $T$ frames, a spatio-temporal ResNet \cite{stafylakis2017combining} is applied. After spatial average-pooling, a feature vector of $T \times 512$ is generated for each video.
\subsection{Setup}

\subsubsection{\textbf{Implementation Details}}
The overall framework is stacked by multiple transformer layers. The layer numbers of each module are specified in Table \ref{table of parameter}.
\begin{table}[h]
\centering
\caption{The layer number of transformers in each module.}
\begin{tabular}{|c|c|c|c|c|}
\hline$\theta_{a}$ & $\theta_{v}$ & $\theta_{j}$ & $\theta_{d-FCD}$ & $\theta_{d-TAM}$ \\
\hline6             & 6             & 6             & 1   & 2  \\
\hline

\end{tabular}

\label{table of parameter}
\end{table}

The pertraining is based on the approach of DeepAVSR \cite{afouras2018deep}. We select the TM-CTC model trained on LRS2~\cite{son2017lip} dataset and use the publicly available, pretrained model.\footnote{\url{https://github.com/lordmartian/deep_avsr}} We adopt modality dropout during training, where dropout is applied to mask the full features of one modality, and the loss for the corresponding labels are also masked. We train the model with a batch size of 12 and Adam optimization~\cite{kingma2014adam} with a learning rate of 1e-5. Every model is trained for 100 iterations in total.
\subsubsection{\textbf{Metrics}}
The metric of our dual-label experiments is the F1 score. Since each dataset has varying degrees of class imbalances, we use average per-class F1 (CF1) and overall F1 (OF1). Actually, OF1 is preferable for class imbalance since it weighs each sample equally. To take the class size into consideration, we also report weighted-averaged F1 (WF1) scores which can account for the contribution of each class by weighting the number of samples of that given class. The accuracy score (ACC) is also used to evaluate the performance of each single modality detection. For comparative analysis with state-of-the-art methods, we adapt to the binary classification metrics employed in their studies.

\subsection{Audio-Visual Deepfake Datasets}
 
\subsubsection{\textbf{DFDC}} DFDC \cite{dolhansky2019deepfake} is a larger-scale dataset containing 128,154 videos, including 104,500 fake videos. Both audio manipulation and visual manipulation are involved in DFDC.  Due to its large scale, previous works~\cite{zhou2021joint,chugh2020not,mittal2020emotions} did not perform experiments on the entire set of videos. A popular practice is to sample a subset from DFDC. Following prior works \cite{chugh2020not,mittal2020emotions}, we sample the subset of 18,000 videos (85:15 train-test split) from DFDC. DFDC does not contain videos in which only the audio component is fake; only those videos with a manipulated visual component are further selected for audio swapping. Similar to \cite{hosler2021deepfakes}, the specific fake audio label is given by comparing audio tracks of fake videos with those of original videos, which is achieved via hashing the audio file sequence.\footnote{\url{https://www.kaggle.com/datasets/basharallabadi/dfdc-video-audio-labels}}

\subsubsection{\textbf{FakeAVCeleb}}
FakeAVCeleb \cite{khalid2021fakeavceleb}  is a multimodal deepfake dataset tailored for audio-visual deepfake detection, consisting of 500 real videos and 19,500 deepfake videos. Its forgery types align more closely with our scenarios, as it includes deepfakes that are manipulated in audio-only, visual-only, and both audio-visual forms. %
\subsubsection{\textbf{LAV-DF}}
Localized Audio Visual DeepFake (LAV-DF)~\cite{cai2022you} is a multimodal dataset in which local-manipulated contents exist in either the audio or video component (or both). It contains 136,304 videos, including 99,873 fake videos. Unlike the above two datasets, where fake content is throughout the entire video or audio signal, manipulation in LAV-DF is driven by transcript content. The word/words are replaced with their antonym(s) to generate small fake segments of audio or video.

\subsection{Comparative Experiments}
%Currently, there are \sj{few} approaches that tackle audio-visual deepfake detection in a dual-label manner. 
To demonstrate the effectiveness of our method, we conduct comparative experiments with state-of-the-art detectors on audio-visual, visual-only, and audio-only deepfakes, respectively. 
\subsubsection{\textbf{Audio-visual deepfake detection}} In this subsection, we first focus on a comparison with audio-visual deepfake detection methods. This means performing predictions at the entire video level, regardless of what the specific fake modality is. We report the AUC scores in Table \ref{tab:BinaryClass}. To ensure a fair comparison, particularly since different methods select various subsets on DFDC, we employ N-fold Cross-Validation across different subsets and report the mean AUC along with the standard deviation on it. As for FakeAVCeleb, we directly use full data following the splits for train/test from the original datasets. Notably, since the output of our method is the independent probability of each modality, we calculate the predicted authenticity of the entire video by multiplying the probability of the real class from each modality, which is then used to compute AUC scores.

\begin{table}[h]

\centering
\caption{Comparison of \textbf{audio-visual} deepfake detection on DFDC and FakeAVCeleb datasets. Results of some methods are cited from AVoiD-DF~\cite{yang2023avoid} and PVASS~\cite{yu2023pvass}.}

\begin{tabular}{c|c|c|c|c}
\hline \multirow{2}{*}{ Audio-Visual Methods }  & \multicolumn{2}{c}{ FakeAVCeleb } & \multicolumn{2}{|c}{ DFDC } \\
\cline { 2 - 5 } &  ACC & AUC &  ACC & AUC \\
\hline 
 
\hline 

%AVN-J [44] (2021)   & 73.2 & 77.6 & 81.1 & 83.3 \\
MDS~\cite{chugh2020not} (2020)  & 82.8 & 86.5 & 89.8 & 91.6 \\
EmoForen~\cite{mittal2020emotions} (2020)  & 78.1 & 79.8 & 80.6 & 84.4 \\
AVFakeNet~\cite{ilyas2023avfakenet} (2022)  & 78.4 & 83.4 & 82.8 & 86.2 \\
VFD~\cite{cheng2022voice} (2022) & 81.5 & 86.1 & 80.9 & 85.1 \\
BA-TFD~\cite{cai2022you} (2022)  & 80.8 & 84.9 & 79.1 & 84.6 \\
JointAV~\cite{zhou2021joint} (2021) & 82.5  & 83.3 &90.2 &91.9\\

AVoiD-DF~\cite{yang2023avoid} (2023) ) & 83.7 & 89.2 &91.4 & 94.8 \\
AVAD~\cite{feng2023self} (2023)  &94.2 &94.5 &93.2 &96.7 \\
PVASS~\cite{yu2023pvass} (2023) &95.7 &97.3 &$\mathbf{96.3}$ &$\mathbf{98.9}$ \\

\hline Ours & $\mathbf{99.7}$ & $\mathbf{99.9}$ & 91.2$\pm{0.3}$  &  96.9$\pm{0.4}$ \\

\hline
\end{tabular}

\label{tab:BinaryClass}
\end{table}

Table \ref{tab:BinaryClass} shows that our framework outperforms all other methods on the FakeAVCeleb dataset and yields competitive results on DFDC. For DFDC, our AUC scores are comparable to very recent methods like PVASS, although our ACC scores are lower. This is because ACC is a metric that is more sensitive to class imbalance. The audio deepfakes are far fewer than visual deepfakes in DFDC, which makes our framework prone to lower scores when calculating overall video-level accuracy. Unlike DFDC, which is predominantly composed of face-swapped videos, FakeAVCeleb introduces more complex cross-modal forgery patterns by incorporating deepfakes generated by Wav2lip. Specifically, the audio content drives the lip movements, creating a unique type of forgery. Our method is particularly well-suited to combat these kinds of forgeries by effectively modeling cross-modal speech correlations. In contrast, methods like EmoForen, which rely on emotion correlation across modalities, fall short in their effectiveness. Our approach, focusing on speech correlation, proves to be more effective in detecting these sophisticated forgeries.

\begin{table}[h]
\centering
\caption {Comparison on local-manipulated audio-visual deepfakes using AUC metrics. Results of other methods are cited from LAV-DF~\cite{cai2022you}.}

\begin{tabular}
{c|c|c|c}
\hline  \multicolumn{4}{c}{ LAV-DF } \\
\hline MDS \cite{chugh2020not} (2020) & EffiViT~\cite{coccomini2022combining} (2022)& BA-TFD~\cite{cai2022you} (2022) & Ours\\
\hline   82.8 & 96.5 & 99.0 &$\mathbf{99.9}$ \\
\hline
\end{tabular}
\label{tab:BinaryLAV}
\end{table}

In addition to traditional deepfake datasets, we also evaluate the performance of our methods on partially modified deepfakes on LAV-DF. These subtle fake contents are more indistinguishable and imperceptible due to their lesser discrepancy from real content. Detecting this type of localized forgery could be more valuable and challenging in practical application scenarios. As shown in Table \ref{tab:BinaryLAV}, the results on LAV-DF show that our method also outperforms the other compared methods which demonstrates our methods can also combat such novelty deepfakes well. 

\subsubsection{\textbf{Visual-only deepfake detection}}%Comparison with uni-modal methods for visual deepfake detection:} }

In this subsection, we report the performance of binary prediction for visual deepfake detection and compare it with state-of-the-art video-only detection methods. This means our framework is fed only with the video stream as input during the test, corresponding to the scenarios of missing audio modality.

\begin{table}[h]

\caption{Comparison of \textbf{visual-only} deepfake detection on DFDC and FakeAVCeleb datasets.}
%\begin{tabular}{c|c|c|c|c}
\footnotesize
\begin{tabular}{@{}c|c|c|c|c@{}}

\hline \multirow{2}{*}{ Visual-Only Methods }  & \multicolumn{2}{c}{ FakeAVCeleb } & \multicolumn{2}{|c}{ DFDC } \\
\cline { 2 - 5 } &  ACC & AUC &  ACC & AUC \\
\hline MesoNet~\cite{afchar2018mesonet} (2018)  & 57.3 & 60.9 & 71.7 & 75.3 \\
Capsule~\cite{nguyen2019capsule}  (2019)  & 68.8 & 70.9 & 50.2 & 53.3 \\
HeadPose~\cite{yang2019exposing} ( (2019) & 45.6 & 49.2 & 51.4 & 55.9 \\
VA-MLP~\cite{8638330}  (2019)   & 65.0 & 67.1 & 59.3 & 61.9 \\
Xception~\cite{rossler2019faceforensics++}  (2019) & 67.9 & 70.5 & 46.5 & 49.9 \\
LipForensics~\cite{haliassos2021lips}  (2021) & 80.1 & 82.4 & 71.3 & 73.5 \\
%RealForensics ~\cite{haliassos2022leveraging}  (2022) \\
DeFakeHop~\cite{chen2021defakehop} (2021)  & 68.3 & 71.6 & 78.9 & 81.1 \\
CViT~\cite{wodajo2021deepfake}  (2021)  & 69.7 & 71.8 & 62.1 & 63.7 \\
MaDD~\cite{zhao2021multi} (2021)   & 77.6 & 79.3 & 82.5 & 84.8 \\
SLADD~\cite{Chen_2022_CVPR} (2022)   & 70.5 & 72.1 & 73.6 & 75.2 \\
RealForensics~\cite{haliassos2022leveraging} (2022) & 90.1 & 92.3 & 89.6 & 91.5 \\

\hline Ours (Video Only)& $\mathbf{99.6}$ & $\mathbf{99.9}$ & $\mathbf{89.3\pm{0.7}}$ & $\mathbf{96.7\pm{0.4}}$  \\
\hline
\end{tabular}
\label{video-only}
\end{table}
 As shown in Table \ref{video-only}, our framework achieves the best performance on both datasets. Our framework outperforms advanced self-supervised methods like SLADD and RealForensics. RealForensics employs audio in its self-supervised training to learn a natural cross-modal representation. We include its results here due to its visual-only settings at test time. Although it achieves competitive results, it leaves high-level semantic features underutilized, placing its performance behind ours. LipForensics emphasizes the effectiveness of high-level semantic information via clip-wise lipreading, which differs from our frame-wise phoneme prediction approach, resulting in their limited performance. We attribute our superior performance to the leveraging of frame-wise phoneme speech features, which is beneficial in capturing more fine-grained forgery clues. 

\subsubsection{\textbf{Audio-only deepfake detection}}%Comparison with uni-modal methods for audio deepfake detection:} }

In this subsection, we report the performance of binary prediction for audio deepfake detection and compare it with works specifically dedicated to combating audio-only deepfakes.  For these experiments, we exclusively feed the audio stream into our framework to align with the scenarios of missing visual modality. Given that DFDC is primarily utilized for visual deepfake detection in current literature, here we only report the performance on FakeAVCeleb. Following their conventions, we use EER (Equal Error Rate) and the AUC score as evaluation metrics. We choose to benchmark against the work presented in DAD-SV~\cite{pianese2022deepfake}, which detects audio deepfakes via the Speaker Verification task. This is because both our methods and DAD-SV share a similar strategy: leveraging external pre-tasks to extract high-level information beneficial for deepfake detection. The key difference is that DAD-SV compares the audio to be detected with pristine audio from the same individual to determine authenticity, requiring them to rely on pre-acquired identity information of the test audio. In contrast, our method focuses more on semantic information for detection.

\begin{table}[h]
\centering
\caption{Comparison of \textbf{audio-only} deepfake detection on FakeAVCeleb dataset.}
\begin{tabular}{lcc}
\hline
Audio-Only Methods & EER & AUC \\
\hline
LCNN-LSTM-sum-p2s~\cite{Wang2021ACS} & 0.39 & 67.3 \\
RawGATstMul~\cite{tak2021end} & 0.66 & 30.5 \\
RawNet2~\cite{tak2021endrawnet2} & 0.55 & 45.5 \\
\hline
ClovaAI~\cite{Chung2020InDO} & 0.40 & 62.8 \\
H/ASP ~\cite{heo2020clova} & 0.14 & 93.5 \\
ECAPA-TDNN~\cite{desplanques2020ecapa} & 0.21 & 86.2 \\
POI-Forensics~\cite{Cozzolino_2023_CVPR} & 0.21 & 85.8 \\
POI-Forensics + aug~\cite{Cozzolino_2023_CVPR} & 0.26 & 82.4 \\
\hline
Ours (Audio Only) & $\mathbf{0.01} $& $\mathbf{99.9}$ \\
\hline
\end{tabular}
\label{tab:audio detection}
\end{table}

 As shown in Table \ref{tab:audio detection}, we compare our results with supervised learning baseline models ~\cite{Wang2021ACS,tak2021end,tak2021endrawnet2} and the speaker verification based methods~\cite{Chung2020InDO,heo2020clova,desplanques2020ecapa,Cozzolino_2023_CVPR}  that DAD-SV employed in their works. Our method demonstrates superior performance, surpassing all baseline methods as well as the best-performing speaker verification method, H/ASP, by more than 6\% on AUC. The results also indicate that both speaker verification methods and our method generally outperform supervised methods. This suggests that the incorporation of external prior information can indeed benefit deepfake detection. Nevertheless, methods that rely on identity information may be constricted by the specific identity in the reference dataset. Our approach focuses more on semantic aspects, allowing our method to achieve further improvement.

\begin{table*}[t]
\centering

\caption{Benefit of AVSR. }
\begin{tabular}{c|c|c|c|c|c|c|c|c|c}
\hline Dataset  &\multicolumn{3}{c|}{DFDC}  &\multicolumn{3}{c|}{LAV-DF}  &\multicolumn{3}{c}{FakeAVCeleb}   \\  \cline{1-10}
\hline Method  & OF1    & CF1   & WF1   & OF1    & CF1   &WF1    & OF1    & CF1   &WF1   \\ \cline{1-10}
\hline w/o AVSR & 77.52  & 75.60 & 76.78 & 95.65 &95.66  &95.53  &98.95 &99.04 &98.70 \\ \cline{1-10}

% above line is origin paper 70 step
\hline w/  AVSR     & 90.24  & 87.80 & 90.24 & 99.85 &99.85  &99.86 &99.87 &99.89 &99.89  \\ \cline{1-10}
\end{tabular}

\label{tab:AVSR}
\end{table*}
\begin{figure*}[t]
 \centering
\includegraphics[width=0.9\textwidth]{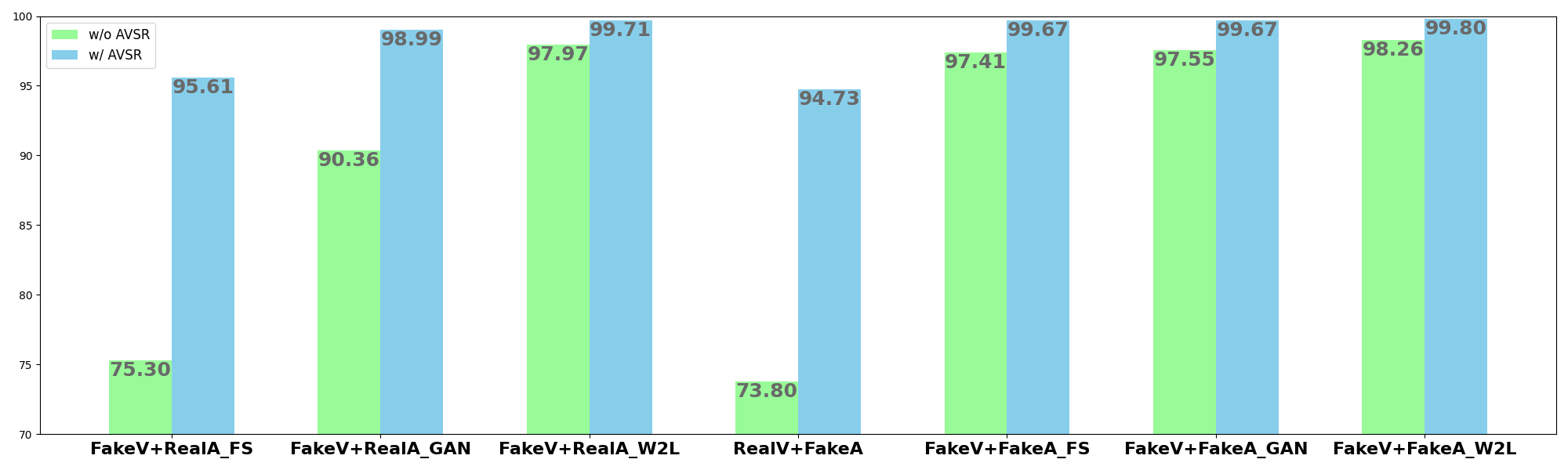} % Reduce the figure size so that it is slightly narrower than the column.
 \caption{The performance comparison on different types of fake videos on the FakeAVCeleb dataset.}
 \label{fig:performance}
\end{figure*}

\subsection{Ablation Study}
\subsubsection{\textbf{Benefit of AVSR}}
To demonstrate the effectiveness of audio-visual speech correlation for dual-label deepfake detection, we train a framework the same as our final framework from scratch without AVSR weights loaded for comparison. 

As shown in Table~\ref{tab:AVSR}, introducing the AVSR task significantly improves the detection performance, which verifies that potential speech correlation across modalities is highly beneficial to synergistic audio-visual deepfake detection. The improvement is not only reflected in globally forged DFDC and FakeAVCeleb but also in locally forged LAV-DF. Albeit LAV-DF proves that changing a few uttering words can even lead to a subtle forgery beyond the perception of humans, leveraging AVSR allows us to extract phoneme-level features frame by frame. The promising potential of these fine-grained semantic features will show more value in real-life scenarios.

%\subsubsection{\textbf{Performance analysis on different fake types}}
The extent of speech correlation might not be uniform and varies based on the specific deepfake generation method employed. To further investigate the impact of potential speech correlation across various deepfake categories, we examine the performance of each fake type on the FakeAVCeleb dataset. As shown in Figure~\ref{fig:performance}, we present a comparative analysis based on the OF1 score for each type, both with and without the incorporation of audio-visual speech recognition. Here we append a suffix behind the forgery type of each modality to denote the manipulated methods, resulting a seven categories, including (1)``FakeV+RealA\_FS": real audio with fake video by FaceSwap;
(2) ``FakeV+RealA\_GAN": real audio with fake video by FSGAN;
(3)``FakeV+RealA\_W2L": real audio with fake video by Wav2Lip;
(4) ``RealV+FakeA": real video with fake audio by SV2TTS;
(5) ``FakeV+FakeA\_FS": fake video by FaceSwap and Wav2Lip, and fake audio by SV2TTS;
(6) ``FakeV+FakeA\_GAN": fake video by FSGAN and Wav2Lip, and fake audio by SV2TTS;
(7) ``FakeV+FakeA\_Ori": fake video by Wav2Lip, and fake audio by SV2TTS. As all types with fake audio are generated by SV2TTS, here we omit it in the suffix. Notably, ``FakeV+FakeA\_Ori" denotes that the fake video content is created solely through the lip-syncing process via Wav2lip, which drives the fake audio to generate lip movements on the original face. Here, we use ``Ori" to denote the absence of any extra face-swapping process.

% Use \bibliography{yourbibfile} instead or the References section will not appear in your paper

As illustrated in Figure \ref{fig:performance}, our method consistently performs well across all types of forgeries. It shows an especially significant improvement in the ``RealV+FakeA" category. This is because the fake audio in this case is generated by the text-driving method and fails to reproduce natural temporal synchronization with the lip movements of the original video. Yet, our method is particularly adept at capturing this audio-visual asynchrony. Moreover, our method surpasses 99\% on all forgery types generated by Wav2Lip, which re-proves our strength in capturing cross-modal fake clues.

\subsubsection{\textbf{Benefit of modality compensation}}
\label{subsection MC}
In this section, we evaluate how the form of the modality compensation strategy affects the performance of dual-label deepfake detection on DFDC. We evaluate the variants of (1) the simple concatenation of two modalities with no compensation (\textbf{None}), as preceding speech recognition does; (2) audio modality compensated (\textbf{Audio}); (3) video modality compensated (\textbf{Video}). We also show the detection performance of each single modality (\textbf{AF1} for audio, and \textbf{VF1} for video).
\begin{table}[h]

\caption{Ablation results of different strategies of modality compensation.}
\centering
\begin{tabular}{c|c|c|c|c|c}

\hline {Compensation}  & OF1 & CF1 & VF1 & AF1 & WF1  \\

\hline None & $88.96$ & $86.02$ & $89.62$ & $82.40$ & $88.57$  \\
\hline Video & $87.79$ & $84.41$ & $88.50$ & $80.17$ & $87.57$ \\
%\hline Audio &\textbf{91.89} & \textbf{88.34} & \textbf{92.69} & \textbf{83.96} & \textbf{91.66}   \\
\hline Audio &\textbf{90.24} & \textbf{87.80} & \textbf{90.84} & \textbf{84.61} & \textbf{90.24}   \\
% above line is origin model 70 step
\hline
\end{tabular}

\label{tab:MCA}
\end{table}
As shown in Table~\ref{tab:MCA}, consistent with our speculation in section \ref{methodMCA}, adopting audio modality compensation yields the most desirable performance. Notably, compared with video compensation, audio compensation improves not only audio detection performance but also that of video. Since the influence of audio on model decision-making tends to be weaker, we believe a balance between audio and visual modalities is achieved by this mechanism and helps both modalities learn better features.

\begin{table*}[t]
\centering
\caption{Generalization study on the influence of different components in Dual-Label Classifier. The CrossFakeAVCeleb denotes trained on DFDC and tested on FakeAVCeleb, and vice versa.}
\begin{tabular}{c|c|c|c|c|c|c|c|c|c|c}
\hline {Methods} & \multicolumn{5}{c|}{ CrossFakeAVCeleb } & \multicolumn{5}{c}{ CrossDFDC } \\
\hline Variants  & OF1 & CF1 & WF1 & VACC & AACC & OF1 & CF1 & WF1 & VACC & AACC \\
\hline w/o TAM\&FCD & $65.77$ & $54.07$ & $57.12$ & $74.73$ & $49.34$ & $55.89$ & $29.23$ & $53.00$ & $58.46$ & $\mathbf{94.65}$ \\
\hline w/o TAM & $67.50$ & $55.53$ & $\mathbf{59.68}$ & $76.41$ & $\mathbf{49.55}$ & $58.28$ & $32.67$ & $55.77$ & $59.19$ & $94.61$ \\

\hline Ours & $\mathbf{68.57}$ & $\mathbf{55.83}$ & $58.77$ & $\mathbf{80.00}$ & $48.89$ & $\mathbf{58.57}$ & $\mathbf{3 6 . 9 7}$ & $\mathbf{5 6 . 7 5}$ & $\mathbf{6 1 . 4 2}$ & $94.07$ \\
\hline
\end{tabular}

\label{tab:generalize}
\end{table*}

\begin{table*}[t]
\centering
\caption{Comparison of our dual-label framework with unimodal approaches and multi-classifier approaches under three test modality-agnostic cases: prediction with visual-only (shown in pink), audio-only (shown in green), or audio-visual input. The best results are highlighted in bold.}
\begin{tabular}{c|c|c|c|c|c|c|c|c|c|c}
%\rowcolors{1}{green!20}{yellow!50}
\hline  \multicolumn{2}{c}{Available Modality} & \multicolumn{1}{|c|}{Classification} & \multicolumn{4}{c|}{ FakeAVCeleb } & \multicolumn{4}{c}{ DFDC } \\
\hline Audio & Visual  &Type & VACC & AACC & VF1 & AF1 & VACC & AACC & VF1 & AF1 \\
\hline$\sqrt{ }$ & -  & Binary & - & 96.99 & - & 97.23 & - & 75.79 & - & 68.04 \\
- & $\sqrt{ }$  & Binary & 89.59 & - & 98.96 & - & 84.19 & - & 84.61 & - \\

%\rowcolor{green!15}
\hline $\sqrt{ }$ & - & Dual-classifier & - & \cellcolor{green!15}97.16 & - & \cellcolor{green!15}97.35 & - & \cellcolor{green!15}97.54 & - & \cellcolor{green!15}66.67 \\

- & $\sqrt{ }$  & Dual-classifier & \cellcolor{red!15}99.50 & - & \cellcolor{red!15}99.74 & - & \cellcolor{red!15}87.96 & - & \cellcolor{red!15}87.95 & - \\
$\sqrt{ }$ & $\sqrt{ }$  & Dual-classifier & 99.64 & 99.88 & 99.81 & 99.88 & 90.07 & 98.34 & 89.33 & 83.14 \\

\hline$\sqrt{ }$ & -  & Triple-classifier & - &\cellcolor{green!15}98.19 & - &\cellcolor{green!15}98.24 & - &\cellcolor{green!15}95.80 & - &\cellcolor{green!15}59.77 \\

- & $\sqrt{ }$  & Triple-classifier & \cellcolor{red!15}99.48 & - &\cellcolor{red!15}99.72 & - & \cellcolor{red!15}87.58 & - & \cellcolor{red!15}87.97 & - \\
$\sqrt{ }$ & $\sqrt{ }$  & Triple-classifier & 99.64 & \textbf{99.95} & 99.81 & \textbf{99.95} & 89.96 & 98.15 & 89.21 & 82.22 \\

\hline$\sqrt{ }$ & -  & Dual-Label & - & \cellcolor{green!15}\textbf{99.95} & - & \cellcolor{green!15}\textbf{99.95} & - &\cellcolor{green!15}98.27 & - & \cellcolor{green!15}81.33 \\

- & $\sqrt{ }$  & Dual-Label & \cellcolor{red!15}99.65 & - & \cellcolor{red!15}99.81 & - & \cellcolor{red!15}90.50 & - & \cellcolor{red!15}89.96 & - \\
$\sqrt{ }$ & $\sqrt{ }$ & Dual-Label & \textbf{99.70} & 99.93 & \textbf{99.84} & 99.93 & \textbf{91.23} & \textbf{98.46} & \textbf{90.83} & \textbf{84.61} \\
%FakeAVCeleb step20:/projects/yucai2/code/deep_avsr-master/audio_visual/NoneDropAblation/DualLabel_FakeAVCeleb_global/DropModality_A/finetune_checkpoints/FakeAVCelebFakeAVCeleball.txt
%DFDCstep30:xin
\hline
\end{tabular}

\label{tab:multimodal}
\end{table*} 

\subsubsection{\textbf{Generalization ability of sub-modules in dual-label classifier}} We expect the design of TAM and FCD in the Dual-Label Classifier to provide a more generalizable audio-visual representation. Here, we develop two variants and conduct a series of experiments of cross-dataset tests to investigate the influence of different components. We train our framework on both FakeAVCeleb and DFDC and mutually perform cross-dataset tests.

As shown in Table~\ref{tab:generalize}, the general results indicate that the absence of the TAM leads to a decline in generalization performance on both datasets. Moreover, the deletion of the FCD further exacerbates this decline, which verifies the importance of bridging the forgery intensity gap across modalities. Overall, performance on both datasets reaches its peak when all components are employed in the Dual-Label Classifier.

\subsubsection{\textbf{Benefit of dual-label classifier}}
To investigate the impact of dual-label manner on the detection of each modality, we adapt our framework into various variants based on the involved modality and classification approach. First, we trained two separate unimodal models for audio and video modalities to serve as baselines. Additionally, we adapted our framework to align with previous multimodal approaches, transforming it into two variants of multi-classifier methods: a dual-classifier (i.e., for audio and visual streams, respectively) as done in AVFakeNet, and a triple-classifier (i.e., for audio, visual, and entire video, respectively) as implemented in JointAV. As presented in Table~\ref{tab:multimodal}, these comparative experiments, along with our proposed method, constitute three groups of experiments. Under each group of experiments, we report three types of test results to examine the performance under modality-agnostic scenarios, i.e., after training is completed in a multimodal setting, the model performs prediction with visual-only (shown in pink), audio-only (shown in green), or audio-visual input during the testing phase.

It can be observed that no matter whether compared to unimodal approaches or classifier variants, our method performs best on all metrics for both datasets, particularly in cases where both audio and video are available. In fact, with the support of AVSR, under such a test case that is consistent with training conditions, the results produced by different classifier variants exhibit minor differences and all surpass unimodal methods. This highlights the advantage of AVSR in modeling cross-modal forgery clues when compared to unimodal methods. 

In more challenging missing modality situations, our method further extends its lead over other variants by maintaining performance on par with audio-visual detection. When either modality is unavailable, the variation in our method's performance on the FakeAVCeleb dataset remains minimal, within decimal point ranges. However, both dual and triple classifiers experience a decline of over 1 percentage point in the audio-only input situation. This gap is more evident in DFDC which is dominated by visual forgeries, where both the AF1 for dual classifiers and triple classifiers exhibit a more significant decline. For video-only detection, our method continues to outperform the unimodal baseline and other variants. The combined results from both datasets show that our method consistently maintains its advantages over unimodal approaches in their respective modalities, regardless of whether the modalities are present in pairs or not.

We attribute the effectiveness of our method to the fact that, compared to multiple classifiers that create a branched structure, the dual-label approach results in a highly-multiplexed unified structure that more effectively preserves the forgery patterns extracted by AVSR. In essence, the dual-label technique optimizes the benefits of AVSR. Overall, the performance on audio-only, visual-only, and audio-visual detection collectively demonstrates that the proposed framework excels in all three fake-modality-agnostic scenarios.

\begin{figure}[h]
 \centering
\includegraphics[width=1.0\columnwidth]{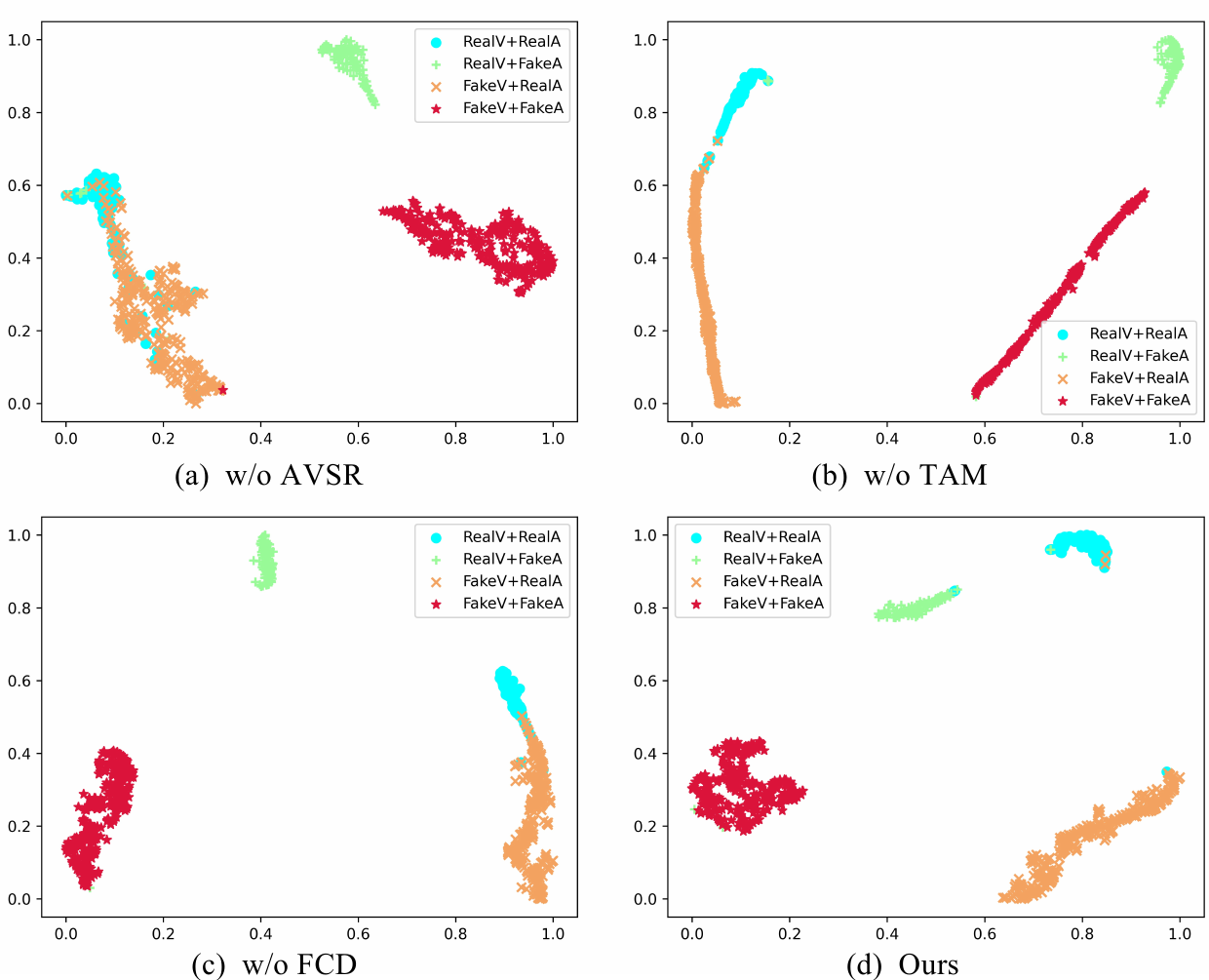} % Reduce the figure size so that it is slightly narrower than the column.
 \caption{T-SNE visualization of feature space learned by variants of our framework.}
 \label{fig:tsnemap}
\end{figure}
\subsection{Visualization}
%\subsubsection{\textbf{Feature Space Learned by Different Variants of The Framework}}

To demonstrate the efficacy of each design, we visualize the feature cluster of four categories with regard to manipulated modalities, namely, ``RealV+RealA", ``RealV+FakeA", ``FakeV+RealA", and ``FakeV+FakeA". We employ t-SNE \cite{van2008visualizing} to visualize the feature distribution of the test set from FakeAVCeleb. As shown in Figure~\ref{fig:tsnemap}, without finetuning on AVSR (Figure a), the model suffers from the most severe feature entanglement. After finetuning on AVSR (Figure b, c, d), the features exhibit better distinguishability. However, the absence of either TAM or FCD still leaves the distinguishability of video-only fakes (orange) not satisfactory enough. Collectively, our framework generates the most favorable feature distribution. As depicted in Figure~\ref{fig:tsnemap}d, the distributions of four categories are disentangled well. Notably, among the three fake distributions, the feature clusters of audio-only fakes (green) and video-only fakes (orange) are relatively close to that of the reals (blue), and the audio-visual fakes (red) are most distant from the reals as they possess more fake compositions, which is consistent with our description of forgery intensity. We believe this optimal feature distribution also reflects the contribution of FCD.

\section{Conclusion}
This paper presents a unified, modality-agnostic framework capable of detecting audio-visual deepfakes, regardless of the number or type of modality involved. We designed a dual-label detection classifier to specifically identify whether either the audio or visual modality (or both) has been manipulated, while effectively handling scenarios with missing modalities. The AVSR is introduced to provide high-level speech correlations across modalities, enhancing the framework's ability to model cross-modal forgery clues. The integration of AVSR with the Dual-Label Classifier results in excellent performance for each individual modality. The overall component design leads our framework to not only outperform various state-of-the-art competitors but also effectively detect modality-agnostic audio-visual deepfakes with promising performance. In our future work, we will expand our investigation to cover a broader spectrum of multimodal forgery types and enhance the performance for complex talking scenarios.

\bibliographystyle{bibtex/IEEEtran}
\bibliography{bibtex/IEEEexample}

\begin{thebibliography}{10}
\providecommand{\url}[1]{#1}
\csname url@samestyle\endcsname
\providecommand{\newblock}{\relax}
\providecommand{\bibinfo}[2]{#2}
\providecommand{\BIBentrySTDinterwordspacing}{\spaceskip=0pt\relax}
\providecommand{\BIBentryALTinterwordstretchfactor}{4}
\providecommand{\BIBentryALTinterwordspacing}{\spaceskip=\fontdimen2\font plus
\BIBentryALTinterwordstretchfactor\fontdimen3\font minus
  \fontdimen4\font\relax}
\providecommand{\BIBforeignlanguage}[2]{{%
\expandafter\ifx\csname l@#1\endcsname\relax
\typeout{** WARNING: IEEEtran.bst: No hyphenation pattern has been}%
\typeout{** loaded for the language `#1'. Using the pattern for}%
\typeout{** the default language instead.}%
\else
\language=\csname l@#1\endcsname
\fi
#2}}
\providecommand{\BIBdecl}{\relax}
\BIBdecl

\bibitem{korshunova2017fast}
I.~Korshunova, W.~Shi, J.~Dambre, and L.~Theis, ``Fast face-swap using
  convolutional neural networks,'' in \emph{Proceedings of the IEEE
  international conference on computer vision}, 2017, pp. 3677--3685.

\bibitem{thies2016face2face}
J.~Thies, M.~Zollhofer, M.~Stamminger, C.~Theobalt, and M.~Nie{\ss}ner,
  ``Face2face: Real-time face capture and reenactment of rgb videos,'' in
  \emph{Proceedings of the IEEE conference on computer vision and pattern
  recognition}, 2016, pp. 2387--2395.

\bibitem{siarohin2019first}
A.~Siarohin, S.~Lathuili{\`e}re, S.~Tulyakov, E.~Ricci, and N.~Sebe, ``First
  order motion model for image animation,'' \emph{Advances in Neural
  Information Processing Systems}, vol.~32, 2019.

\bibitem{ping2018clarinet}
W.~Ping, K.~Peng, and J.~Chen, ``Clarinet: Parallel wave generation in
  end-to-end text-to-speech,'' \emph{arXiv preprint arXiv:1807.07281}, 2018.

\bibitem{kameoka2018non}
H.~Kameoka, T.~Kaneko, K.~Tanaka, and N.~S.-V. Hojo, ``Non-parallel
  many-to-many voice conversion using star generative adversarial networks,''
  in \emph{SLT Workshop}, 2018, pp. 18--21.

\bibitem{prajwal2020lip}
K.~Prajwal, R.~Mukhopadhyay, V.~P. Namboodiri, and C.~Jawahar, ``A lip sync
  expert is all you need for speech to lip generation in the wild,'' in
  \emph{MM}, 2020, pp. 484--492.

\bibitem{cheng2022videoretalking}
K.~Cheng, X.~Cun, Y.~Zhang, M.~Xia, F.~Yin, M.~Zhu, X.~Wang, J.~Wang, and
  N.~Wang, ``Videoretalking: Audio-based lip synchronization for talking head
  video editing in the wild,'' 2022.

\bibitem{chugh2020not}
K.~Chugh, P.~Gupta, A.~Dhall, and R.~Subramanian, ``Not made for each
  other-audio-visual dissonance-based deepfake detection and localization,'' in
  \emph{MM}, 2020, pp. 439--447.

\bibitem{gu2020deepfake}
Y.~Gu, X.~Zhao, C.~Gong, and X.~Yi, ``Deepfake video detection using
  audio-visual consistency,'' in \emph{International Workshop on Digital
  Watermarking}.\hskip 1em plus 0.5em minus 0.4em\relax Springer, 2020, pp.
  168--180.

\bibitem{mittal2020emotions}
T.~Mittal, U.~Bhattacharya, R.~Chandra, A.~Bera, and D.~Manocha, ``Emotions
  don't lie: An audio-visual deepfake detection method using affective cues,''
  in \emph{MM}, 2020, pp. 2823--2832.

\bibitem{zhou2021joint}
Y.~Zhou and S.-N. Lim, ``Joint audio-visual deepfake detection,'' in
  \emph{ICCV}, 2021, pp. 14\,800--14\,809.

\bibitem{yang2023avoid}
W.~Yang, X.~Zhou, Z.~Chen, B.~Guo, Z.~Ba, Z.~Xia, X.~Cao, and K.~Ren,
  ``Avoid-df: Audio-visual joint learning for detecting deepfake,'' \emph{IEEE
  Transactions on Information Forensics and Security}, vol.~18, pp. 2015--2029,
  2023.

\bibitem{cheng2022voice}
H.~Cheng, Y.~Guo, T.~Wang, Q.~Li, T.~Ye, and L.~Nie, ``Voice-face homogeneity
  tells deepfake,'' \emph{arXiv preprint arXiv:2203.02195}, 2022.

\bibitem{haliassos2022leveraging}
A.~Haliassos, R.~Mira, S.~Petridis, and M.~Pantic, ``Leveraging real talking
  faces via self-supervision for robust forgery detection,'' in \emph{CVPR},
  2022, pp. 14\,950--14\,962.

\bibitem{cai2022you}
Z.~Cai, K.~Stefanov, A.~Dhall, and M.~Hayat, ``Do you really mean that? content
  driven audio-visual deepfake dataset and multimodal method for temporal
  forgery localization,'' \emph{arXiv preprint arXiv:2204.06228}, 2022.

\bibitem{ilyas2023avfakenet}
H.~Ilyas, A.~Javed, and K.~M. Malik, ``Avfakenet: A unified end-to-end dense
  swin transformer deep learning model for audio--visual deepfakes detection,''
  \emph{Applied Soft Computing}, vol. 136, p. 110124, 2023.

\bibitem{feng2023self}
C.~Feng, Z.~Chen, and A.~Owens, ``Self-supervised video forensics by
  audio-visual anomaly detection,'' \emph{arXiv preprint arXiv:2301.01767},
  2023.

\bibitem{yu2023pvass}
Y.~Yu, X.~Liu, R.~Ni, S.~Yang, Y.~Zhao, and A.~C. Kot, ``Pvass-mdd: Predictive
  visual-audio alignment self-supervision for multimodal deepfake detection,''
  \emph{IEEE Transactions on Circuits and Systems for Video Technology}, 2023.

\bibitem{li2020face}
L.~Li, J.~Bao, T.~Zhang, H.~Yang, D.~Chen, F.~Wen, and B.~Guo, ``Face x-ray for
  more general face forgery detection,'' in \emph{CVPR}, 2020, pp. 5001--5010.

\bibitem{shiohara2022detecting}
K.~Shiohara and T.~Yamasaki, ``Detecting deepfakes with self-blended images,''
  in \emph{CVPR}, 2022, pp. 18\,720--18\,729.

\bibitem{gu2022exploiting}
Q.~Gu, S.~Chen, T.~Yao, Y.~Chen, S.~Ding, and R.~Yi, ``Exploiting fine-grained
  face forgery clues via progressive enhancement learning,'' in \emph{AAAI},
  vol.~36, no.~1, 2022, pp. 735--743.

\bibitem{yu2022focus}
C.~Yu, P.~Chen, J.~Dai, X.~Wang, W.~Zhang, J.~Liu, and J.~Han, ``Focus by
  prior: Deepfake detection based on prior-attention,'' in \emph{2022 IEEE
  International Conference on Multimedia and Expo (ICME)}.\hskip 1em plus 0.5em
  minus 0.4em\relax IEEE, 2022, pp. 1--6.

\bibitem{zhao2021multi}
H.~Zhao, W.~Zhou, D.~Chen, T.~Wei, W.~Zhang, and N.~Yu, ``Multi-attentional
  deepfake detection,'' in \emph{CVPR}, 2021, pp. 2185--2194.

\bibitem{agarwal2019protecting}
S.~Agarwal, H.~Farid, Y.~Gu, M.~He, K.~Nagano, and H.~Li, ``Protecting world
  leaders against deep fakes.'' in \emph{CVPR Workshops}, vol.~1, 2019, p.~38.

\bibitem{sun2021improving}
Z.~Sun, Y.~Han, Z.~Hua, N.~Ruan, and W.~Jia, ``Improving the efficiency and
  robustness of deepfakes detection through precise geometric features,'' in
  \emph{CVPR}, 2021, pp. 3609--3618.

\bibitem{li2018eye}
Y.~Li, M.-C. Chang, and S.~Lyu, ``In ictu oculi: Exposing ai created fake
  videos by detecting eye blinking,'' in \emph{WIFS}.\hskip 1em plus 0.5em
  minus 0.4em\relax IEEE, 2018, pp. 1--7.

\bibitem{gu2022delving}
\BIBentryALTinterwordspacing
Z.~Gu, Y.~Chen, T.~Yao, S.~Ding, J.~Li, and L.~Ma, ``Delving into the local:
  Dynamic inconsistency learning for deepfake video detection,'' in
  \emph{AAAI}.\hskip 1em plus 0.5em minus 0.4em\relax {AAAI} Press, 2022, pp.
  744--752. [Online]. Available:
  \url{https://ojs.aaai.org/index.php/AAAI/article/view/19955}
\BIBentrySTDinterwordspacing

\bibitem{zhang2021detecting}
D.~Zhang, C.~Li, F.~Lin, D.~Zeng, and S.~Ge, ``Detecting deepfake videos with
  temporal dropout 3dcnn.'' in \emph{IJCAI}, 2021, pp. 1288--1294.

\bibitem{gu2021spatiotemporal}
Z.~Gu, Y.~Chen, T.~Yao, S.~Ding, J.~Li, F.~Huang, and L.~Ma, ``Spatiotemporal
  inconsistency learning for deepfake video detection,'' in \emph{CVPR}, 2021,
  pp. 3473--3481.

\bibitem{haliassos2021lips}
A.~Haliassos, K.~Vougioukas, S.~Petridis, and M.~Pantic, ``Lips don't lie: A
  generalisable and robust approach to face forgery detection,'' in
  \emph{CVPR}, 2021, pp. 5039--5049.

\bibitem{martin2022vicomtech}
J.~M. Mart{\'\i}n-Do{\~n}as and A.~{\'A}lvarez, ``The vicomtech audio deepfake
  detection system based on wav2vec2 for the 2022 add challenge,'' in
  \emph{ICASSP}.\hskip 1em plus 0.5em minus 0.4em\relax IEEE, 2022, pp.
  9241--9245.

\bibitem{xie2021siamese}
Y.~Xie, Z.~Zhang, and Y.~Yang, ``Siamese network with wav2vec feature for
  spoofing speech detection.'' in \emph{Interspeech}, 2021, pp. 4269--4273.

\bibitem{wang2021investigating}
X.~Wang and J.~Yamagishi, ``Investigating self-supervised front ends for speech
  spoofing countermeasures,'' \emph{arXiv preprint arXiv:2111.07725}, 2021.

\bibitem{baevski2020wav2vec}
A.~Baevski, Y.~Zhou, A.~Mohamed, and M.~Auli, ``wav2vec 2.0: A framework for
  self-supervised learning of speech representations,'' \emph{NeurIPS},
  vol.~33, pp. 12\,449--12\,460, 2020.

\bibitem{pianese2022deepfake}
A.~Pianese, D.~Cozzolino, G.~Poggi, and L.~Verdoliva, ``Deepfake audio
  detection by speaker verification,'' in \emph{2022 IEEE International
  Workshop on Information Forensics and Security (WIFS)}.\hskip 1em plus 0.5em
  minus 0.4em\relax IEEE, 2022, pp. 1--6.

\bibitem{hosler2021deepfakes}
B.~Hosler, D.~Salvi, A.~Murray, F.~Antonacci, P.~Bestagini, S.~Tubaro, and
  M.~C. Stamm, ``Do deepfakes feel emotions? a semantic approach to detecting
  deepfakes via emotional inconsistencies,'' in \emph{CVPR}, 2021, pp.
  1013--1022.

\bibitem{shi2021learning}
B.~Shi, W.-N. Hsu, K.~Lakhotia, and A.~Mohamed, ``Learning audio-visual speech
  representation by masked multimodal cluster prediction,'' in \emph{ICLR},
  2021.

\bibitem{campbell2008processing}
R.~Campbell, ``The processing of audio-visual speech: empirical and neural
  bases,'' \emph{Philosophical Transactions of the Royal Society B: Biological
  Sciences}, vol. 363, no. 1493, pp. 1001--1010, 2008.

\bibitem{almajai2007maximising}
I.~Almajai and B.~Milner, ``Maximising audio-visual speech correlation.'' in
  \emph{AVSP}, 2007, p.~17.

\bibitem{9151013}
S.~Agarwal, H.~Farid, O.~Fried, and M.~Agrawala, ``Detecting deep-fake videos
  from phoneme-viseme mismatches,'' in \emph{CVPR Workshops}, 2020, pp.
  2814--2822.

\bibitem{ma2021smil}
M.~Ma, J.~Ren, L.~Zhao, S.~Tulyakov, C.~Wu, and X.~Peng, ``Smil: Multimodal
  learning with severely missing modality,'' in \emph{Proceedings of the AAAI
  Conference on Artificial Intelligence}, vol.~35, no.~3, 2021, pp. 2302--2310.

\bibitem{7952625}
S.~Petridis, Z.~Li, and M.~Pantic, ``End-to-end visual speech recognition with
  lstms,'' in \emph{2017 IEEE International Conference on Acoustics, Speech and
  Signal Processing (ICASSP)}, 2017, pp. 2592--2596.

\bibitem{afouras2018deep}
T.~Afouras, J.~S. Chung, A.~Senior, O.~Vinyals, and A.~Zisserman, ``Deep
  audio-visual speech recognition,'' \emph{TPAMI}, 2018.

\bibitem{8682566}
S.~Zhang, M.~Lei, B.~Ma, and L.~Xie, ``Robust audio-visual speech recognition
  using bimodal dfsmn with multi-condition training and dropout
  regularization,'' in \emph{ICASSP 2019 - 2019 IEEE International Conference
  on Acoustics, Speech and Signal Processing (ICASSP)}, 2019, pp. 6570--6574.

\bibitem{9004036}
T.~Makino, H.~Liao, Y.~Assael, B.~Shillingford, B.~Garcia, O.~Braga, and
  O.~Siohan, ``Recurrent neural network transducer for audio-visual speech
  recognition,'' in \emph{2019 IEEE Automatic Speech Recognition and
  Understanding Workshop (ASRU)}, 2019, pp. 905--912.

\bibitem{graves2006connectionist}
A.~Graves, S.~Fern{\'a}ndez, F.~Gomez, and J.~Schmidhuber, ``Connectionist
  temporal classification: labelling unsegmented sequence data with recurrent
  neural networks,'' in \emph{ICML}, 2006, pp. 369--376.

\bibitem{ma2021end}
P.~Ma, S.~Petridis, and M.~Pantic, ``End-to-end audio-visual speech recognition
  with conformers,'' in \emph{ICASSP}.\hskip 1em plus 0.5em minus 0.4em\relax
  IEEE, 2021, pp. 7613--7617.

\bibitem{dolhansky2019deepfake}
B.~Dolhansky, R.~Howes, B.~Pflaum, N.~Baram, and C.~C. Ferrer, ``The deepfake
  detection challenge (dfdc) preview dataset,'' \emph{arXiv preprint
  arXiv:1910.08854}, 2019.

\bibitem{cheng2021mltr}
X.~Cheng, H.~Lin, X.~Wu, F.~Yang, D.~Shen, Z.~Wang, N.~Shi, and H.~Liu, ``Mltr:
  Multi-label classification with transformer,'' \emph{arXiv preprint
  arXiv:2106.06195}, 2021.

\bibitem{vaswani2017attention}
A.~Vaswani, N.~Shazeer, N.~Parmar, J.~Uszkoreit, L.~Jones, A.~N. Gomez,
  {\L}.~Kaiser, and I.~Polosukhin, ``Attention is all you need,'' in
  \emph{NeurIPS}, 2017, pp. 5998--6008.

\bibitem{zhang2017s3fd}
S.~Zhang, X.~Zhu, Z.~Lei, H.~Shi, X.~Wang, and S.~Z. Li, ``S3fd: Single shot
  scale-invariant face detector,'' in \emph{ICCV}, 2017, pp. 192--201.

\bibitem{stafylakis2017combining}
T.~Stafylakis and G.~Tzimiropoulos, ``Combining residual networks with lstms
  for lipreading,'' \emph{Interspeech}, pp. 3652--3656, 2017.

\bibitem{son2017lip}
J.~Son~Chung, A.~Senior, O.~Vinyals, and A.~Zisserman, ``Lip reading sentences
  in the wild,'' in \emph{Proceedings of the IEEE conference on computer vision
  and pattern recognition}, 2017, pp. 6447--6456.

\bibitem{kingma2014adam}
D.~P. Kingma and J.~Ba, ``Adam: A method for stochastic optimization,''
  \emph{arXiv preprint arXiv:1412.6980}, 2014.

\bibitem{khalid2021fakeavceleb}
\BIBentryALTinterwordspacing
H.~Khalid, S.~Tariq, M.~Kim, and S.~S. Woo, ``Fake{AVC}eleb: A novel
  audio-video multimodal deepfake dataset,'' in \emph{NeurIPS}, 2021. [Online].
  Available: \url{https://openreview.net/forum?id=TAXFsg6ZaOl}
\BIBentrySTDinterwordspacing

\bibitem{coccomini2022combining}
D.~A. Coccomini, N.~Messina, C.~Gennaro, and F.~Falchi, ``Combining
  efficientnet and vision transformers for video deepfake detection,'' in
  \emph{International Conference on Image Analysis and Processing}.\hskip 1em
  plus 0.5em minus 0.4em\relax Springer, 2022, pp. 219--229.

\bibitem{afchar2018mesonet}
D.~Afchar, V.~Nozick, J.~Yamagishi, and I.~Echizen, ``Mesonet: a compact facial
  video forgery detection network,'' in \emph{WIFS}.\hskip 1em plus 0.5em minus
  0.4em\relax IEEE, 2018, pp. 1--7.

\bibitem{nguyen2019capsule}
H.~H. Nguyen, J.~Yamagishi, and I.~Echizen, ``Capsule-forensics: Using capsule
  networks to detect forged images and videos,'' in \emph{ICASSP}.\hskip 1em
  plus 0.5em minus 0.4em\relax IEEE, 2019, pp. 2307--2311.

\bibitem{yang2019exposing}
X.~Yang, Y.~Li, and S.~Lyu, ``Exposing deep fakes using inconsistent head
  poses,'' in \emph{ICASSP}.\hskip 1em plus 0.5em minus 0.4em\relax IEEE, 2019,
  pp. 8261--8265.

\bibitem{8638330}
F.~Matern, C.~Riess, and M.~Stamminger, ``Exploiting visual artifacts to expose
  deepfakes and face manipulations,'' in \emph{WACV Workshops}, 2019, pp.
  83--92.

\bibitem{rossler2019faceforensics++}
A.~Rossler, D.~Cozzolino, L.~Verdoliva, C.~Riess, J.~Thies, and M.~Nie{\ss}ner,
  ``Faceforensics++: Learning to detect manipulated facial images,'' in
  \emph{ICCV}, 2019, pp. 1--11.

\bibitem{chen2021defakehop}
H.-S. Chen, M.~Rouhsedaghat, H.~Ghani, S.~Hu, S.~You, and C.-C.~J. Kuo,
  ``Defakehop: A light-weight high-performance deepfake detector,'' in
  \emph{2021 IEEE International conference on Multimedia and Expo
  (ICME)}.\hskip 1em plus 0.5em minus 0.4em\relax IEEE, 2021, pp. 1--6.

\bibitem{wodajo2021deepfake}
D.~Wodajo and S.~Atnafu, ``Deepfake video detection using convolutional vision
  transformer,'' \emph{arXiv preprint arXiv:2102.11126}, 2021.

\bibitem{Chen_2022_CVPR}
L.~Chen, Y.~Zhang, Y.~Song, L.~Liu, and J.~Wang, ``Self-supervised learning of
  adversarial example: Towards good generalizations for deepfake detection,''
  in \emph{Proceedings of the IEEE/CVF Conference on Computer Vision and
  Pattern Recognition (CVPR)}, June 2022, pp. 18\,710--18\,719.

\bibitem{Wang2021ACS}
X.~Wang and J.~Yamagishi, ``A comparative study on recent neural spoofing
  countermeasures for synthetic speech detection,'' in \emph{Interspeech},
  2021, pp. 4259--4263.

\bibitem{tak2021end}
H.~Tak, J.-w. Jung, J.~Patino, M.~Kamble, M.~Todisco, and N.~Evans,
  ``End-to-end spectro-temporal graph attention networks for speaker
  verification anti-spoofing and speech deepfake detection,'' \emph{Automatic
  Speaker Verification and Spoofing Countermeasures Challenge}, 2021.

\bibitem{tak2021endrawnet2}
H.~Tak, J.~Patino, M.~Todisco, A.~Nautsch, N.~Evans, and A.~Larcher,
  ``End-to-end anti-spoofing with rawnet2,'' in \emph{ICASSP 2021-2021 IEEE
  International Conference on Acoustics, Speech and Signal Processing
  (ICASSP)}.\hskip 1em plus 0.5em minus 0.4em\relax IEEE, 2021, pp. 6369--6373.

\bibitem{Chung2020InDO}
J.~S. Chung, J.~Huh, S.~Mun, M.~Lee, H.-S. Heo, S.~Choe, C.~Ham, S.-Y. Jung,
  B.-J. Lee, and I.~Han, ``In defence of metric learning for speaker
  recognition,'' in \emph{Interspeech}, 2020, pp. 2977--2981.

\bibitem{heo2020clova}
H.~S. Heo, B.-J. Lee, J.~Huh, and J.~S. Chung, ``Clova baseline system for the
  voxceleb speaker recognition challenge 2020,'' \emph{arXiv preprint
  arXiv:2009.14153}, 2020.

\bibitem{desplanques2020ecapa}
B.~Desplanques, J.~Thienpondt, and K.~Demuynck, ``Ecapa-tdnn: Emphasized
  channel attention, propagation and aggregation in tdnn based speaker
  verification,'' pp. 3830--3834, 2020.

\bibitem{Cozzolino_2023_CVPR}
D.~Cozzolino, A.~Pianese, M.~Nie{\ss}ner, and L.~Verdoliva, ``Audio-visual
  person-of-interest deepfake detection,'' in \emph{Proceedings of the IEEE/CVF
  Conference on Computer Vision and Pattern Recognition (CVPR) Workshops}, June
  2023, pp. 943--952.

\bibitem{van2008visualizing}
L.~Van~der Maaten and G.~Hinton, ``Visualizing data using t-sne.''
  \emph{Journal of machine learning research}, vol.~9, no.~11, 2008.

\end{thebibliography}
\end{document}